\documentclass[reprint,aps,pra,twocolumn,showpacs]{revtex4-2}

\usepackage{graphicx}
\usepackage{dcolumn}
\usepackage{bm}
\usepackage{hyperref}

\hypersetup{colorlinks=true, citecolor=blue, urlcolor=blue, linkcolor=blue}
\usepackage{slashed}  

\usepackage{amsmath}
\usepackage{amssymb} 
\usepackage{natbib}
\usepackage{mathrsfs}
\usepackage{flushend}
\makeatletter

\newcommand{\Rmnum}[1]{\expandafter\@slowromancap\romannumeral #1@}
\makeatother
\allowdisplaybreaks[3]

\begin{document}	
\preprint{APS/123-QED}

\title{Active control of the peak value of the Hanbury Brown--Twiss effect using coherent light by lensless holographic projection}

\author{Liming Li}
\email{liliming@sdut.edu.cn}
\author{Yuhan Guo}
\author{Chunguang Meng}
\author{Wenfei Zhang}
\author{Gongxiang Wei}
\author{Huiqiang Liu}
\email{liuhq@sdut.edu.cn}
\affiliation{School of Physics and Optoelectronic Engineering, Shandong University of Technology, Zibo 255049, China\\}
\author{Wei Li}
\affiliation{Hefei National Research Center for Physical Sciences at the Microscale and School of Physical Sciences, University of Science and Technology of China, Hefei, 230026, China\\}
\date{\today}

\begin{abstract}
Computer-generated holography enables projection of target patterns onto designated planes, providing deterministic control over the probability density function of the projected light intensity. Here, we introduce an active control scheme for the peak value of the Hanbury Brown--Twiss effect, $g^{(2)}(0)$, utilizing lensless holographic projection with coherent light. Notably, single-frame holographic projection yields a markedly different $g^{(2)}(0)$ from its multiframe-averaged counterpart due to the presence of coherent speckle noise. With the coherent speckle noise suppression, we derive an analytical expression $g^{(2)}(0)$ on holographic projection plane, revealing that it is determined by the target coherence length, its statistics, and the numerical aperture of projection system. Our experimental results show good agreement with the theoretical analysis, confirming the joint influence of these factors. By employing dynamic sparse target patterns, we achieve a maximum $g^{(2)}(0)$ of $39.77$. Numerical simulations, benchmarked against experimental measurements, reveal that coherent speckle noise enhances $g^{(2)}(0)$ through mutual superposition with the target pattern, leading to a joint modulation of intensity fluctuations. In summary, by manipulating multiple controllable parameters, we establish a robust strategy for tailoring $g^{(2)}(0)$, paving the way for advanced applications in speckle imaging and optical metrology.
\end{abstract}
\maketitle

\section{Introduction}
The Hanbury Brown--Twiss (HBT) effect was first observed in 1956 during the development of stellar intensity interferometry~\cite{Hanbury1956Correlation,Hanbury1956A}. This seminal work introduced the fundamental concept of multi-photon interference to the scientific community~\cite{Scully1997Quantum,Rodney2000TheQuantum,Shih2020AnIntroduction}. From a microscopic perspective, the joint detection probability of two photons at the same spatial position in a thermal field is double that of detecting them at distinct positions~\cite{Scully1997Quantum}, a phenomenon widely known as photon bunching. Macroscopically, this effect arises because the intensity auto-correlation inherently exceeds the cross-correlation between distinct points in the detection plane, driven by the intensity fluctuations characteristic of thermal speckle~\cite{Goodman2007Speckle}. According to statistical optics, the probability density function (PDF) of the light intensity for thermal speckle follows a negative exponential distribution~\cite{Goodman1976Some}. Therefore, the second-order auto-correlation at zero delay, $g^{(2)}(0)$, is exactly twice that of the corresponding cross-correlation.

Photon bunching has found widespread application across optics, astronomy~\cite{KarlMN2022Comparing,CrawfordOE2023Towards,RubioPRR2023Non}, radar and remote sensing~\cite{Omar2016SAHanbury,Liu2021PRLimproved,LiuPRL2025active}, particle physics~\cite{Jeltes2007NatureComparison,LampenPRL2019Lampen,PlumbergPRC2022Hanbury}, and cold atom physics~\cite{Schellekens2005ScienceHanbury,Khakimov2016natureGhost,Observation2022Rosenberg}. Conventional HBT interferometry with (pseudo)thermal light has the peak-to-background ratio at 2:1~\cite{Hanbury1956Correlation,Liu2009Norder}. Superbunching, which exceeds the classical limit, significantly enhances key imaging performance metrics, including visibility~\cite{NiePRA2021Noise,ChenOE2023Fourier,WuCPB2024High,LimingOE2026Integrating} and spatial resolution~\cite{ZhangLP2016HIGH,Kuplicki2016OESPECKLE,ZhouOE2023Ghost,jechowNP2013Enhanced,PascucciNC2019Compress,ChoiNL2022Wide,BenderOp2021Circumventing,VigorenJosaa2018Optical,AffannoukoueOp2023Super,ChaigneOP2016Super,HojmanOE2017Photoacoustic,ZhouOL2024Data,LiuOL2019Label}. This has driven advances in computational ghost imaging~\cite{NiePRA2021Noise,ChenOE2023Fourier,WuCPB2024High,LimingOE2026Integrating,ZhangLP2016HIGH,Kuplicki2016OESPECKLE,ZhouOE2023Ghost} and super-resolution microscopy techniques such as two-photon fluorescence microscopy~\cite{jechowNP2013Enhanced,PascucciNC2019Compress,ChoiNL2022Wide}, random illumination microscopy~\cite{BenderOp2021Circumventing,VigorenJosaa2018Optical,AffannoukoueOp2023Super}, and photoacoustic microscopy~\cite{ChaigneOP2016Super,HojmanOE2017Photoacoustic,ZhouOL2024Data,LiuOL2019Label}. Beyond quantum platforms, including nanolasers~\cite{MarconiPRX2018Far,WangPRA2020Superthermal}, superradiant emitters~\cite{JahnkePRA2016Giant,WangAOM2021Optically}, and tunnel junctions~\cite{ChristopherSA2019Photon,KimPRL2024Superbunched}, superbunching has also been demonstrated using entirely classical architectures. Notable examples include tailored optical materials~\cite{BrombergNP2010Hanbury,DeependraArXiv2025Statistics}, dynamic phase and amplitude modulation~\cite{Hong2012Two,Bromberg2014Generating,Zhang2019Superbunching,Liu2021Generation,Luo2021Two-photon,Chen2024Fourier,Zhou2017Superbunching,Zhou2019Superbunching,Liu2021Liu}, holographic speckle engineering~\cite{AmaralPRA2015Tailoring,BenderOp2018Customizing,HanPRL2023Tailoring}, and polarization interference~\cite{Luo2021Temporal,Ye2022Antibunching,Ding2025Spatial}.

While various methods have been proposed to realize the superbunching effect, directly projecting artificial target patterns offers a simple yet highly efficient approach. A key advantage of our work lies in leveraging these artificial patterns to digitally endow the projected light with reconfigurability. This flexibility allows researchers to precisely control $g^{(2)}(0)$ of the bunching effect without being constrained by physical realizability. Furthermore, the lensless projection of artificial dynamic patterns via optical holography significantly enriches the experimental results of ghost imaging~\cite{LimingOE2026Integrating}. A prominent example~\cite{GuoArXiv2026Generalized} is single-pixel imaging (SPI) based on a digital micromirror device (DMD). Our approach fundamentally enriches the DMD operational paradigm in SPI: it shifts from conventional intensity projection through relay lenses governed by geometrical optics~\cite{Duarte2008Single,Edgar2019Principles,Gibson2020Single,Song2025Advances} to lensless holographic illumination employing a binary-amplitude (0--1) computer-generated holography (CGH) rooted in wave optics.

In this work, we derive analytical expressions for the normalized second-order correlation function $g^{(2)}(0)$ on the projection plane under coherent speckle noise suppression, revealing that this peak value is governed jointly by the target pattern and the projection system. We propose and experimentally validate a practical method for predicting $g^{(2)}(0)$ from a single-frame speckle pattern. Furthermore, we demonstrate active control over the HBT peak value in lensless holographic projection by tuning $g^{(2)}(0)$ and the coherence length of the target pattern, the CGH size, and the number of averaged frames for coherent speckle noise suppression. Finally, we quantify the influence of coherent speckle noise on $g^{(2)}(0)$ at the holographic projection plane.

\section{Theoretical model}
We analytically derive the second-order spatial correlation function at the holographic projection plane under noise-suppressed conditions, employing an incoherent framework with the magnification factor of one (a detailed discussion in Appendix \ref{appendixA}). Given that target patterns are typically digitized in experiments, a discrete model is adopted in theoretical analysis. For simplicity, yet without loss of generality, the derivation is performed for the one-dimensional case; however, extension to the two-dimensional scenario is straightforward. When the $\kappa$th power of a dynamic thermal speckle ${I_\text{T}}({x,t})$ serves as the target pattern, the intensity distribution ${I_\text{P}}({\xi ,t} )$ on the projection plane at position $\xi$ and time $t$ satisfies:
\begin{equation}\label{EQ01}
	{I_\text{P}}( {\xi ,t} )= \sum\limits_{i  = 1}^\mathcal{N} {{I_\text{T}^\kappa}( {{x_i },t} )\cdot{{\rm{somb}}^2}[ {k{N_{\rm{A}}}( {\xi  - {x_i }} )} ]},\\
\end{equation}
where $\mathcal{N}$ is the total number of effective points on the target plane, $\text{somb}(x)=2\text{J}_1(x)/x$ is the sombrero function, $k=2\pi/\lambda$ is the wave vector corresponding to the wavelength $\lambda$, and $N_{\rm{A}}$ is the numerical aperture of the projection system, respectively. Here, $\text{J}_1(x)$ denotes the first-order Bessel function of the first kind. 

Theoretically, the normalized second-order spatial correlation function of the speckle target pattern  ${I_\text{T}^\kappa}( {x,t})$ can be expressed as~\cite{Shih2020AnIntroduction}:
\begin{equation}\label{EQ02}
	g^{(2)}_{\text{T}|\kappa}(x_i, x_j)=1+[g^{(2)}_{\text{T}|\kappa}(0)-1]\cdot \text{somb}^2[k{N}^{\prime}_{\rm{A}}(x_i-x_j)],\\
\end{equation}
where ${N}^{\prime}_{\rm{A}}$ is the numerical aperture of the experimental setup that generates the initial thermal speckle ${I_\text{T}}({x,t})$, and $g^{(2)}_{\text{T}|\kappa}(0)=g^{(2\kappa)}_{\text{ther}}(0)/[g^{(\kappa)}_{\text{ther}}(0)]^2$ represents the degree of second-order coherence of the target pattern ${I_\text{T}^\kappa}( {x,t})$. Here, $g^{(\kappa)}_{\text{ther}}(0)=\kappa!$ denotes the degree of the $\kappa$th-order coherence of the thermal speckle ${I_\text{T}}({x,t})$~\cite{Liu2009Norder} (See Appendix \ref{appendixC} for details). Using computational holography, any $\kappa$th power of speckle intensity can serve as the target pattern. Simple calculation yield values of $4/\pi$, 2 and 6 for $g^{(2)}_{\text{T}|\kappa=0.5}(0)$, $g^{(2)}_{\text{T}|\kappa=1}(0)$ and $g^{(2)}_{\text{T}|\kappa=2}(0)$, respectively. 

Based on Eqs.~(\ref{EQ01})-(\ref{EQ02}) and statistical theory, the second-order spatial correlation function evaluated at coincident points $\xi_1=\xi_2=\xi$ on the projection plane reduces to:
\begin{widetext}
	\begin{equation}\label{EQ03}
		\begin{split}
			g_{\text{P}}^{(2)}(\xi_1=\xi_2=\xi)&=\left \langle{I_{\text{P}}^{2} }( {\xi ,t} )  \right \rangle_t /\left \langle{I_\text{P}}( {\xi ,t} )   \right \rangle_t^{2}\\
			&=1+\frac{\sum\limits_{i = 1}^\mathcal{N} 
				\left({{\rm{somb}}^2}[ {k{N_{\rm{A}}}(\xi-{x_i})}]
				\sum\limits_{j=1}^{\mathcal{N}}[g^{(2)}_{\text{T}|\kappa}(x_i, x_j)-1]\cdot
				{{\rm{somb}}^2}[{k{N_{\rm{A}}}\left({\xi  - {x_j}} \right)}]\right) }
			{   \left(\sum\limits_{i= 1}^\mathcal{N} {{\rm{somb}}^2}\left[ {k{N_{\rm{A}}}\left( {\xi - {x_i}} \right)} \right]\right)^2},\\
		\end{split}
	\end{equation}
\end{widetext}
where $\left\langle{\cdot\cdot\cdot}\right\rangle_t $ denotes the ensemble average over time $t$. The detail deduction is provided in Appendix \ref{appendixD}. Note that $g_{\text{P}}^{(2)}(\xi)$ depends on both the target pattern and the numerical aperture $N_{\rm{A}}$ of projection system. If the boundary of the target pattern extends to infinity, $g_{\text{P}}^{(2)}(\xi)$ becomes independent of the position $\xi_1$ and $\xi_2$ but relies solely on the distance $\xi_1-\xi_2$ between the two detection points (See Eq.~(\ref{app016}) in Appendix \ref{appendixD} for proof). Consequently, $g_{\text{P}}^{(2)}(\xi)$ can be simply written as $g_{\text{P}}^{(2)}(0)$. It is straightforward to show that the maximum value of $g_{\text{P}}^{(2)}(0)$ corresponds to $g^{(2)}_{\text{T}|\kappa}(0)$ in a perfect point-to-point projection system. Consequently, the peak attenuation, defined as $g^{(2)}_{\mathrm{T}|\kappa}(0)-g_{\mathrm{P}}^{(2)}(0)$, increases monotonically as ${{N_{\rm{A}}}}$ of the projection system decreases. 

To obtain the curve of the normalized second-order correlation function of the projection field with dynamic patterns, conventional measurements rely on intensity correlations of multi-frame sampling. However, if we only aim to measure the peak value, single-frame sampling is feasible because the results of the ensemble average over time $t$ and over space $\xi$ are equivalent; i.e., $\left \langle{I_{\text{P}}^{2} }( {\xi ,t} )  \right \rangle_t=\left \langle{I_{\text{P}}^{2} }( {\xi ,t} )  \right \rangle_{\xi}$ and $\left \langle{I_\text{P}}( {\xi ,t} )   \right \rangle_t=\left \langle{I_\text{P}}( {\xi ,t} )   \right \rangle_{\xi}$. This stems from the inherent properties of dynamic patterns: the intensity time series at a detected point will manifest somewhere in the detected plane at an arbitrary instant. Thus, the bunching peak of the projection patten can be estimated as follows:
\begin{equation}\label{EQ04}
	\begin{split}
		g_{\text{P}}^{(2)}(0)=\frac{\mathbf{V}[{I_\text{P}}( {\xi ,t} )]+\mathbf{E}^{2} [ {I_\text{P}}( {\xi ,t} )  ] }{\mathbf{E}^{2} [ {I_\text{P}}( {\xi ,t} )  ]},
	\end{split}
\end{equation}
where the variance $\mathbf{V}[{I_\text{P}}( {\xi ,t})]=\left \langle{I_\text{P}^2}( {\xi ,t})   \right \rangle_{\xi}-\left \langle{I_\text{P}}( {\xi ,t})   \right \rangle_{\xi}^2$ and the expectation $\mathbf{E}[{I_\text{P}}( {\xi ,t})]=\left \langle{I_\text{P}}({\xi ,t})   \right \rangle_{\xi}$ can be measured from a single-frame speckle pattern, respectively.

\section{Experimental verification}
\begin{figure}[!b]
	\centering
	\includegraphics[width=0.42\textwidth]{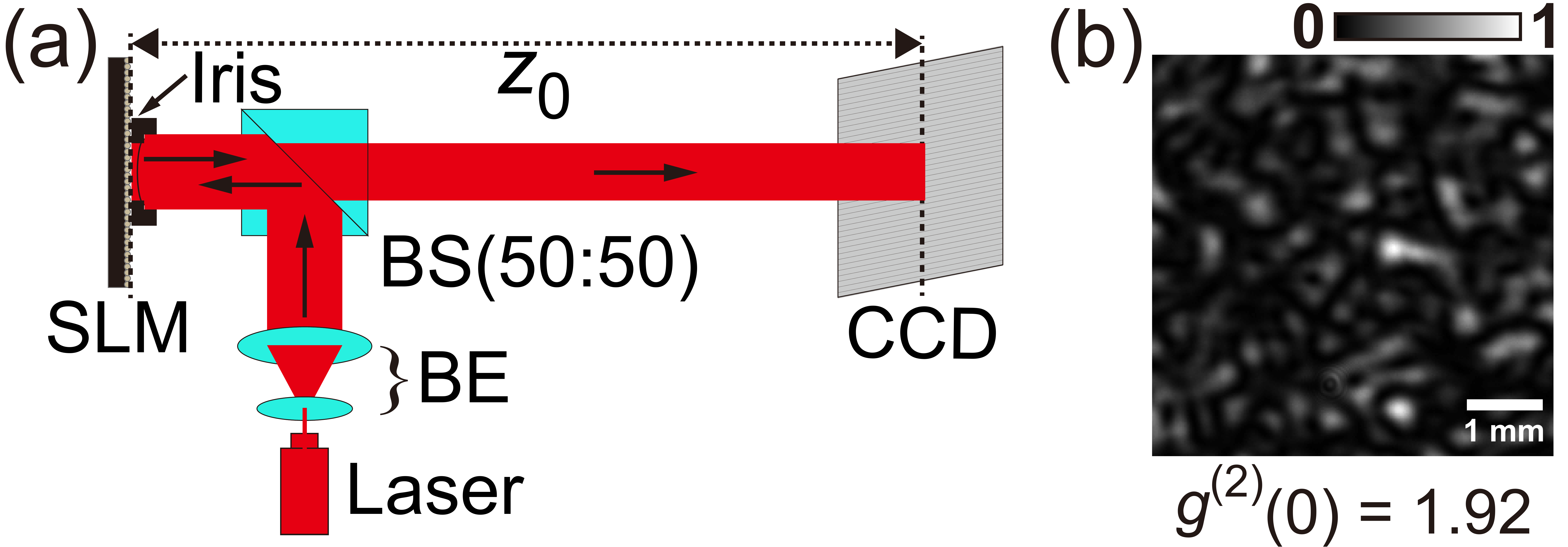}
	\caption{(a) Schematic of the experimental setup for measuring diffraction light intensity from a phase-only spatial light modulator (SLM). BE, beam expander; BS, beam splitter; CCD, charge-coupled device. (b) Single-frame experimental speckle pattern generated by a static chaotic source. The corresponding $g^{(2)}(0)$ is 1.92 calculated via Eq.~(\ref{EQ04}).
	\label{01SchemeAndSpeckle}}
\end{figure}

After the coherent speckle noise is effectively suppressed by multi-frame intensity average, we demonstrate the above theoretical analysis. Figure~\ref{01SchemeAndSpeckle}(a) shows the experimental setup for holographic projection used to measure diffraction intensity from a phase-only spatial light modulator (SLM, pixle size $12.3\times 12.3$  $\mu \text{m}^2$, total pixels $1280\times 1027$, GCI-770401, Daheng Optics, China). A single-mode continuous-wave laser beam with a wavelength $\lambda$ = 632.8 nm was expanded, collimated and reshaped by a beam expander (BE), and then reflected by a 50:50 non-polarized beam splitter (BS). The reflected beam then impinges normally on the active area of the SLM. The phase-encoded laser beam subsequently transmits through the BS and is collected by a charge coupled device (CCD) camera. The distance $z_0$ between the CCD and the SLM is 54.6 cm. The CCD captures far-field diffraction patterns (1.0\,ms exposure).
A lens phase factor $\text{exp}\{-ikr_{\text{SLM}}^2/(2z_0)\}$ establishes a Fourier-transform relation between the SLM and camera planes for lensless projection at $z_{0}$. This superposition is omitted from further discussion for simplicity. Furthermore, we can control the diffraction diameter $D$ of the SLM effective window using an iris, which was placed as close as possible to the SLM.

Actually, target patterns of dynamic Rayleigh speckle sequences can be created in advance by the experimental setup shown in Fig.~\ref{01SchemeAndSpeckle}(a). To create a speckle sequence, dynamic random phase factors $\text{exp}\{i\phi(\vec{r}_{\text{SLM}}, t)\}$ are loaded onto the SLM, where $\phi(\vec{r}_{\text{SLM}}, t)\in [0, 2\pi)$ is a uniformly distributed random phase encoded onto the SLM and completely independent of space and time. Figure~\ref{01SchemeAndSpeckle}(b) shows a single-frame experimental speckle pattern with a diffraction diameter $D_{\rm{chaotic}}=4$ mm of the chaotic source. According to the single-frame evaluation method by using Eq.~(\ref{EQ04}), the $g^{(2)}(0)$ of the experimental speckle in Fig.~\ref{01SchemeAndSpeckle}(b) is 1.92, which is in good agreement with the theoretical results $g^{(2)}_{\text{ther}}(0)=2$. To actively control $g_{\text{P}}^{(2)}(0)$ of the projected pattern, we employ two kinds of target patterns, Rayleigh speckle and sparse patterns, for the experimental investigation.

\subsection{\label{sec:Chaotic} Rayleigh speckle and its functional transformations}
\begin{figure}[!b]
	\centering
	\includegraphics[width=0.49\textwidth]{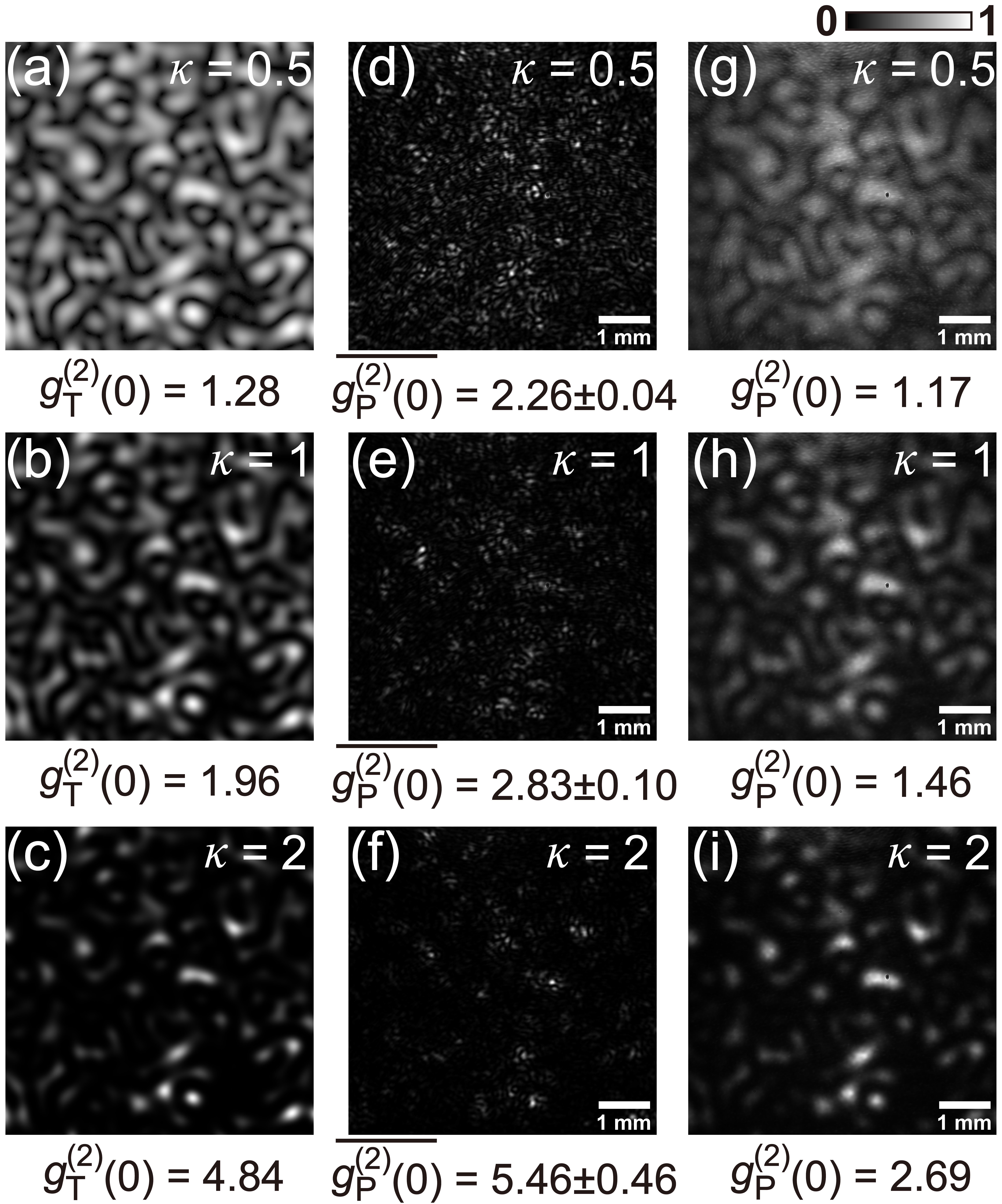}
	\caption{Target patterns of speckle (a)-(c) and holographic projection patterns (d)-(i). Plotted are the $\kappa$th power of the simulated noiseless Rayleigh speckle for $\kappa$ = (a) 0.5, (b) 1 and (c) 2, respectively. (d), (e) and (f) show one of the 1,000 holographic projection results corresponding to the target patterns in (a), (b) and (c), respectively. (g), (h) and (i) show the intensity-averaged patterns of these 1,000 holographic projection results corresponding to (d), (e) and (f), respectively. The evaluation of the peak value by using Eq.~(\ref{EQ04}) is displayed below each subgraph.\label{02ThermalAndHol}}
\end{figure}

Owing to inevitable noise in the experiment, we compute the noiseless target pattern $I_{\text{T}}$ of Rayleigh speckle via numerical simulation. This is achieved by applying the Huygens--Fresnel (HF) principle to the far-field diffraction of a pre-established random phase, using a chaotic source aperture of $D_{\text{chaotic}} = 1\,\text{mm}$~\cite{Goodman1995Introduction}. After applying the $\kappa$th power operation to the Rayleigh speckle ${I_\text{T}}$, figures~\ref{02ThermalAndHol}(a), \ref{02ThermalAndHol}(b) and \ref{02ThermalAndHol}(c) show the target patterns for the cases $\kappa$ = 0.5, 1 and 2 from top to bottom, respectively. With the aid of the Gerchberg–Saxton (GS) algorithm~\cite{ZhouOE2023Ghost,Gerchberg1976A}, we can obtain phase-only CGHs for these target patterns. Since coherent speckle noise cannot be avoided in a single holographic reconstruction~\cite{Chang2015SpeckleAO,Liu2022Double,Zhu2024ThreeD,Dong2025PhaseNC}, we experimentally captured 1,000 projection results using 1,000 frames of CGHs generated by the independent restart GS algorithm with different random initial phases for each individual target pattern. One of these 1,000 results is shown in Figs.~\ref{02ThermalAndHol}(d), \ref{02ThermalAndHol}(e) and \ref{02ThermalAndHol}(f), which correspond to the target patterns in Figs.~\ref{02ThermalAndHol}(a), \ref{02ThermalAndHol}(b) and \ref{02ThermalAndHol}(c), respectively. Here, the effective diameter $D_{\rm{CGH}}$ of CGH is 4 mm. With identical experimental diffraction condition, the fourfold increase in $D_{\rm{CGH}}$ relative to $D_{\rm{chaotic}}$ reduces the reconstructed speckle grain size by a factor of four. Consequently, the single-frame holographic projection yields an intensity distribution jointly determined by the target pattern and coherent diffraction noise. Meanwhile, the intensity-averaged patterns of these 1,000 projection results are shown in Figs.~\ref{02ThermalAndHol}(g), \ref{02ThermalAndHol}(h) and \ref{02ThermalAndHol}(i), which correspond to the single-frame reconstruction patterns in Figs.~\ref{02ThermalAndHol}(d), \ref{02ThermalAndHol}(e) and \ref{02ThermalAndHol}(f), respectively. It is obvious that the coherent speckle noise in Figs.~\ref{02ThermalAndHol}(d), \ref{02ThermalAndHol}(e) and \ref{02ThermalAndHol}(f) introduced by the coherence of the laser beam, can be greatly suppressed by the averaging operation.

Here, the $g_{\text{T}}^{(2)}(0)$ for these target patterns in Figs.~\ref{02ThermalAndHol}(a), \ref{02ThermalAndHol}(b) and \ref{02ThermalAndHol}(c) are 1.28, 1.96 and 4.84, respectively. These results closely align with the theoretical values $g^{(2)}_{\text{T}|\kappa}(0)$. For these single-frame projections, we evaluate the performance using the averaged peak value $\overline{g_{\text{P}}^{(2)}(0)}$ across all recorded patterns. The average peak values with standard deviation of the 1,000 projection results are $2.26\pm0.04$, $2.83\pm0.10$ and $5.46\pm0.46$, which correspond to the target patterns in Figs.~\ref{02ThermalAndHol}(a), \ref{02ThermalAndHol}(b) and \ref{02ThermalAndHol}(c), respectively. Interestingly, these projection results demonstrate the super-bunching characteristics, even though the corresponding target patterns ($\kappa=$ 0.5 and $\kappa=$ 1) without it. In addition, the $g_{\text{P}}^{(2)}(0)$ for those intensity-averaged patterns in Figs.~\ref{02ThermalAndHol}(g), \ref{02ThermalAndHol}(h) and \ref{02ThermalAndHol}(i) are 1.17, 1.46 and 2.69, respectively. As expected by Eq.~(\ref{EQ03}), those peak values $g^{(2)}_{\text{P}}(0)$ of the intensity averages of holographic projection patterns are not greater than the corresponding peak values $g^{(2)}_{\text{T}}(0)$ of the target patterns. While coherent speckle noise is generally detrimental to holographic reconstruction, it proves beneficial for boosting $g^{(2)}_{\text{P}}(0)$ in our scheme. Furthermore, after the squaring operation in the case $\kappa=$ 2, the target pattern in Fig.~\ref{02ThermalAndHol}(c) has a broader dynamic range of light intensity, so the peak values of projection results in Figs.~\ref{02ThermalAndHol}(f) and \ref{02ThermalAndHol}(i) are significantly degraded by environmental noise. 

\begin{figure}[!b]
	\centering
	\includegraphics[width=0.49\textwidth]{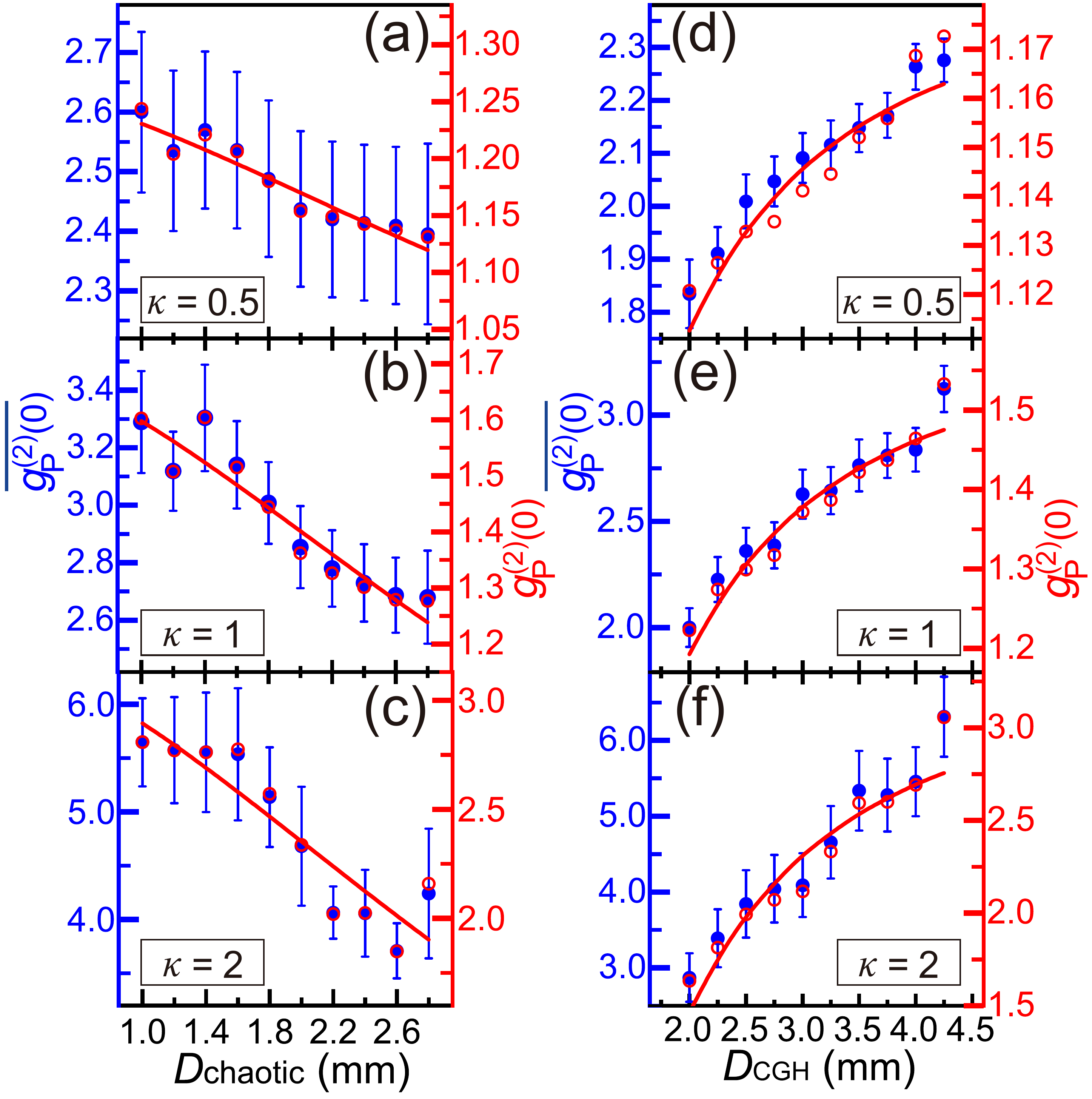}
	\caption{Relationships between $g_{\text{P}}^{(2)}(0)$ and the chaotic source size $D_{\rm{chaotic}}$ for the cases $\kappa$ = (a) 0.5, (b) 1, and (c) 2 from top to bottom. Relationships between $g_{\text{P}}^{(2)}(0)$ and the CGH size $D_{\rm{CGH}}$ for the cases $\kappa$ = (d) 0.5, (e) 1, and (f) 2 from top to bottom. The blue solid circles represent the average peak value $\overline{g_{\text{P}}^{(2)}(0)}$ calculated from 1,000 projection pattern sequences. The error bars (standard deviation) are depicted as blue short line segments. The red hollow circles denote the $g_{\text{P}}^{(2)}(0)$ values for the intensity-averaged patterns of these 1,000-frame projection results. The red solid curves represent the theoretical fits obtained by using Eq.~(\ref{EQ03}).\label{03FunctionByThermal}}
\end{figure}

Besides increasing the degree of second-order coherence function $g^{(2)}_{\text{T}|\kappa}(0)$ of the target pattern, equation ~(\ref{EQ03}) indicates two other methods to enhance $g_{\text{P}}^{(2)}(0)$: increasing the coherence length of the target pattern and improving the numerical aperture of the projection system. It is well known that the coherence length of speckle is inversely proportional to the size of chaotic source~\cite{Goodman2007Speckle}. By varying the size of the chaotic source in numerical simulation, a series of target patterns exhibiting Rayleigh speckle can be generated based on the HF principle. With the CGH size maintained at $D_{\rm{CGH}}$ = 4 mm, the corresponding holographic projection patterns were obtained (these are not displayed here; only the $g_{\text{P}}^{(2)}(0)$ estimated by using Eq.~(\ref{EQ04}) are presented). Figures~\ref{03FunctionByThermal}(a), \ref{03FunctionByThermal}(b), and \ref{03FunctionByThermal}(c) show the relationships between $g_{\text{P}}^{(2)}(0)$ and the chaotic source size $D_{\rm{chaotic}}$ for $\kappa$ = 0.5, 1, and 2, respectively. The blue solid circles represent the average peak value $\overline{g_{\text{P}}^{(2)}(0)}$ calculated from 1,000 projection sequences for a single target pattern with error bars (standard deviation) depicted as short blue line segments. The red hollow circles denote the $g_{\text{P}}^{(2)}(0)$ derived from the intensity average of these 1,000 projection results. The red solid curves represent the theoretical fits obtained by using Eq.~(\ref{EQ03}). Notably, a larger $D_{\rm{chaotic}}$ yields a shorter coherence length, leading to a decrease in $g_{\text{P}}^{(2)}(0)$.

Similarly, figures~\ref{03FunctionByThermal}(d), \ref{03FunctionByThermal}(e), and \ref{03FunctionByThermal}(f) illustrate the relationships between $g_{\text{P}}^{(2)}(0)$ and the CGH size $D_{\rm{CGH}}$ for $\kappa$ = 0.5, 1, and 2, respectively. In this case, the chaotic source size $D_{\rm{chaotic}}$ generating the target speckle patterns is fixed at 1 mm. Here the data in Figs.~\ref{03FunctionByThermal}(d)-\ref{03FunctionByThermal}(f) are presented in a manner consistent with those in Figs.~\ref{03FunctionByThermal}(a)-\ref{03FunctionByThermal}(c), no further description is provided. It is well established that, for a fixed diffraction distance between the CGH plane and the projection plane, a larger $D_{\rm{CGH}}$ corresponds to a larger numerical aperture in the holographic setup~\cite{goodman2005introductionholography} resulting in a higher $g_{\text{P}}^{(2)}(0)$ in the projection plane. 

Notably, $g_{\text{P}}^{(2)}(0)$ obtained under single-frame and multi-frame averaging modes exhibits a strong linear correlation, with Pearson correlation coefficient (PCC)~\cite{Diouf2022Demon} consistently exceeding 0.97 across all six subfigures in Figs.~\ref{02ThermalAndHol}. This indicates that Eq.~(\ref{EQ03}) can predict single-frame $g_{\mathrm{P}}^{(2)}(0)$ using only two known values as input, thereby eliminating the need to account for coherent speckle noise. What's more, it's crucial to distinguish the role of the pinhole diameter $D$, which serves as the control parameter for both $D_{\rm{chaotic}}$ (in numerical simulations) and $D_{\rm{CGH}}$ (in experimental measurement). In Figs.~\ref{03FunctionByThermal}(a)-\ref{03FunctionByThermal}(c), $D_{\rm{chaotic}}$ is varied to control the spatial coherence length of the intensity in target plane, while $D_{\rm{CGH}}$ remains constant. Conversely, in Figs.~\ref{03FunctionByThermal}(d)-\ref{03FunctionByThermal}(f), $D_{\rm{chaotic}}$ is kept constant, while $D_{\rm{CGH}}$ is varied discretely to control the numerical aperture of holographic projection. Consequently, they affect $g_{\text{P}}^{(2)}(0)$ of the projection pattern in opposite ways, resulting in the two distinctly different trends shown in Fig.~\ref{03FunctionByThermal}.

\begin{figure}[t!]
	\centering
	\includegraphics[width=0.49\textwidth]{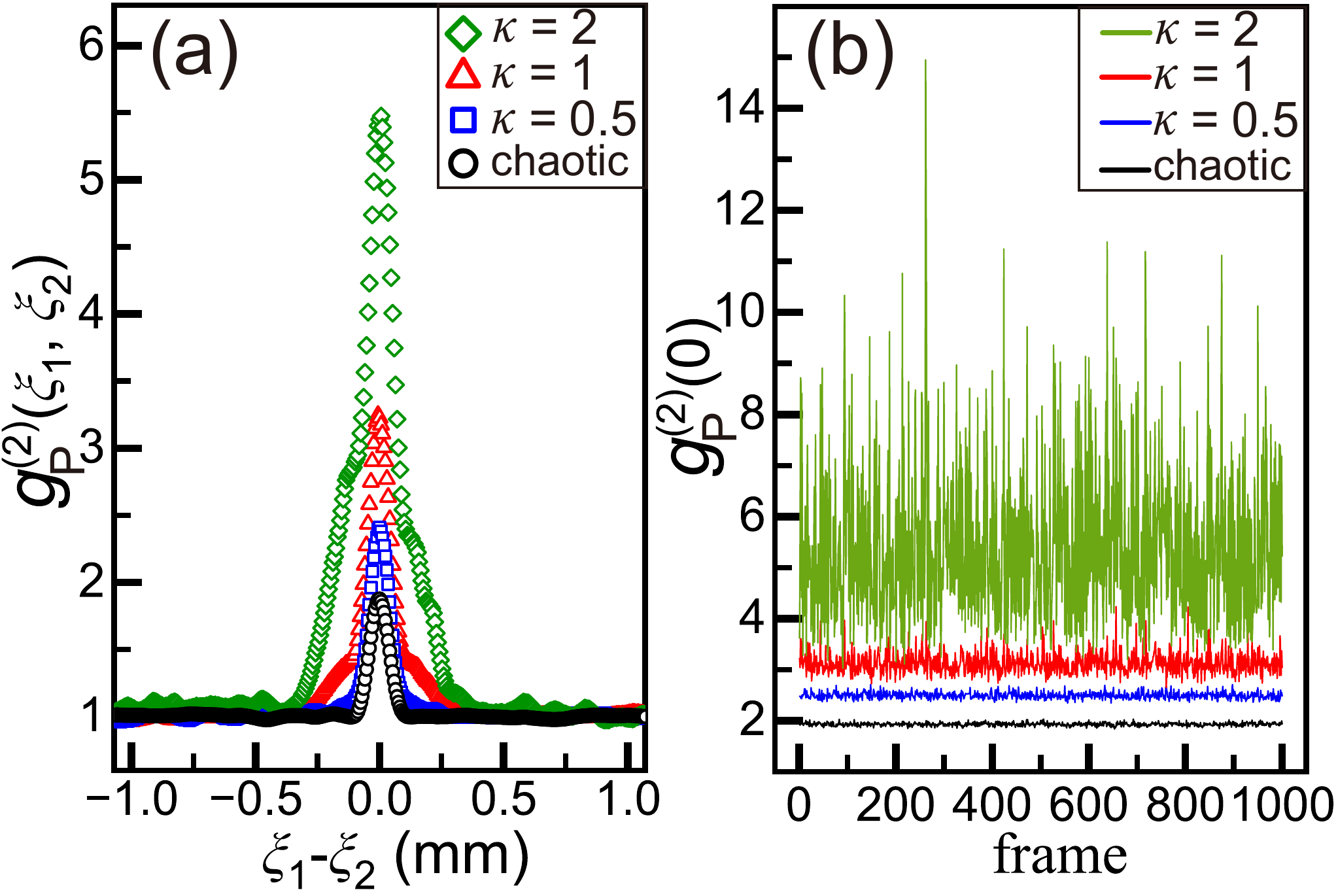}
	\caption{Two measurement methods for $g_{\text{P}}^{(2)}(0)$: (a) intensity correlation of multi-frame patterns and (b) statistical analysis of single-frame patterns. (a) The second-order bunching effect curves calculated from 10,000 frames of Rayleigh speckles (black hollow circles) and holographic projections for $\kappa$ = 0.5, 1, and 2 (blue hollow squares, red hollow triangles, and green hollow diamonds, respectively). (b) Dynamic curves of $g_{\text{P}}^{(2)}(0)$ measured by Eq.~(\ref{EQ04}) from 1,000 frames of Rayleigh speckles (black line) and holographic projections corresponding to $\kappa$ = 0.5, 1, and 2 (blue, red, and green lines, respectively).\label{04Thermalg2bySingleorMulti}}
\end{figure}

To further verify the single-frame measurement method described by Eq.~(\ref{EQ04}), we experimentally captured 10,000 frames of holographic projection patterns generated from distinct simulated noiseless Rayleigh target speckle sequences and measured $g_{\text{P}}^{(2)}(0)$ via the intensity correlation of multi-frames patterns. Additionally, $g^{(2)}(0)$ of a dynamic speckle from a chaotic source under the identical experimental setup is analyzed for comparison. Figure~\ref{04Thermalg2bySingleorMulti} compares the $g_{\text{P}}^{(2)}(0)$ results obtained by two methods: intensity correlation using multi-frame patterns and statistical analysis using single-frame patterns. Figure~\ref{04Thermalg2bySingleorMulti}(a) displays the experimental results of the bunching effect derived from multi-frame intensity correlation for the chaotic source (black hollow circles) and three types of CGHs corresponding to speckle sequences of the chaotic source after power transformation with $\kappa$ = 0.5 (blue hollow squares), 1 (red hollow triangles), and 2 (green hollow diamonds). As expected, the $g_{\text{P}}^{(2)}(0)$ value for the bunching effect of the chaotic source is 1.88, which is close to theoretical value of $g^{(2)}_{\text{ther}}(0)=2$. Notably, the three types of holographic projected patterns exhibit the super-bunching effect with peak values of 2.41, 3.24 and 5.45, corresponding to $\kappa$ = 0.5, 1, and 2, respectively.

Furthermore, the dynamic curves of $g_{\text{P}}^{(2)}(0)$ estimated via statistical analysis of single-frame patterns by using Eq.~(\ref{EQ04}) are shown in Fig.~\ref{04Thermalg2bySingleorMulti}(b). These curves cover the first 1,000 frames of the Rayleigh speckle sequence (black line) and the holographic projection patterns corresponding to $\kappa$ = 0.5, 1, and 2 (blue, red, and green lines, respectively). Since the data points on the curves in Fig.~\ref{04Thermalg2bySingleorMulti}(b) are derived from distinct target patterns, the statistical analysis based on single-frame patterns is influenced by the specific target pattern itself. Here, the average peak values $\overline{g_{\text{P}}^{(2)}(0)}$ with standard deviation calculated over 10,000 frames for the Rayleigh speckle and the holographic projection cases with $\kappa$ = 0.5, 1, and 2, are $1.93\pm0.03$, $2.47\pm0.07$, $3.12\pm0.20$ and $5.33\pm1.50$, respectively. It is evident that the peak value results measured by these two methods are in close agreement. The above results show that a higher $g^{(2)}_{\text{T}}(0)$ of the target pattern yields a higher $g^{(2)}_{\text{P}}(0)$ of the projection pattern. We now investigate holographic projection using artificially designed 0-1 binary sparse patterns.

\subsection{\label{sec:Sparse} Sparse pattern}
\begin{figure}[!t]
	\centering
	\includegraphics[width=0.49\textwidth]{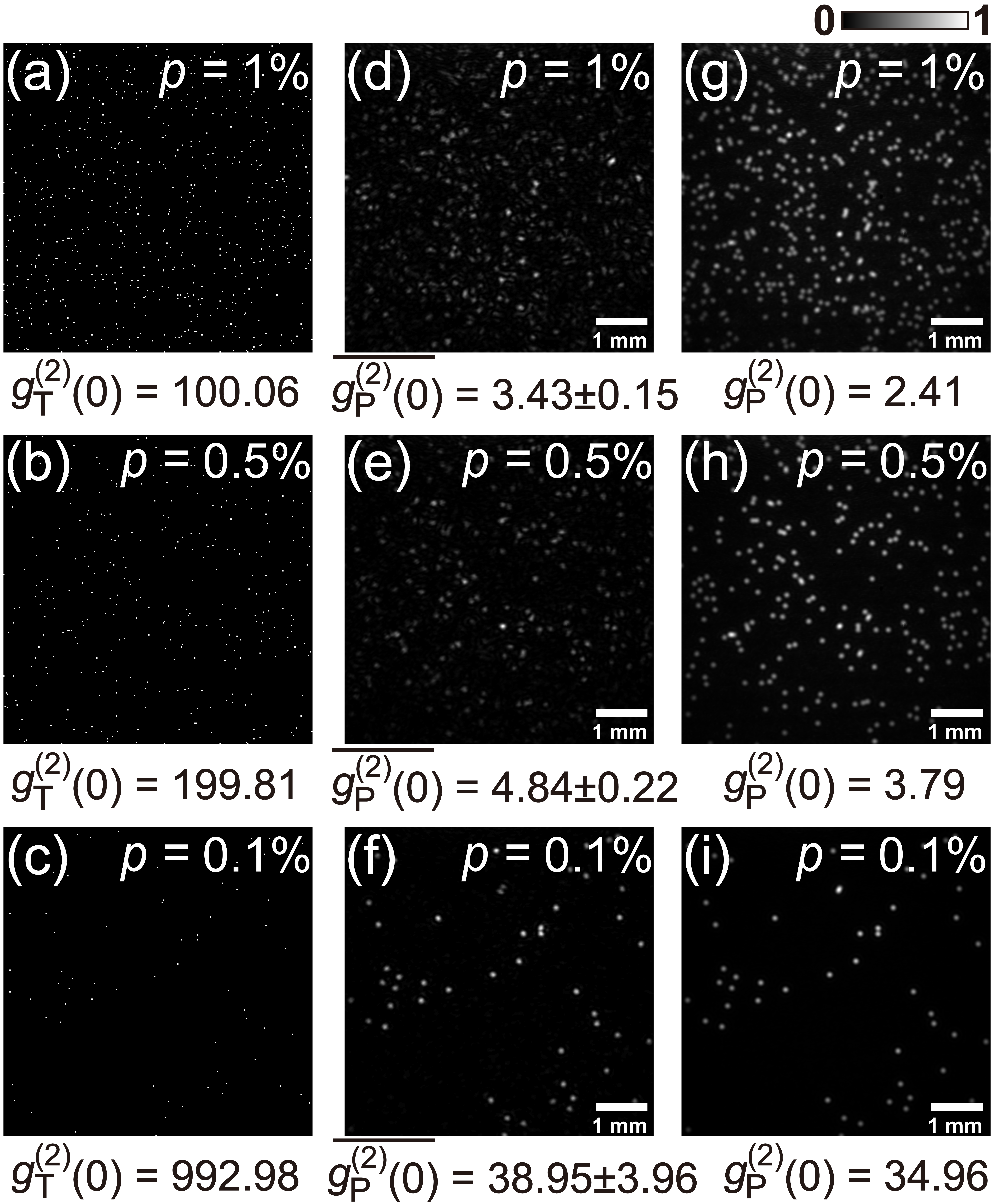}
	\caption{0-1 binary sparse target patterns (a)-(c) and corresponding holographic projection patterns (d)-(i). The sparse patterns were generated from an all-zeros matrix by randomly inserting 1 with a proportion $p$ of (a) 1$\%$, (b) 0.5$\%$ and (c) 0.1$\%$, respectively. (d), (e) and (f) show single instances selected from the 100 projection results corresponding to the target patterns in (a), (b) and (c), respectively. (g), (h) and (i) display the intensity averages of these 100 holographic projection results corresponding to (d), (e) and (f), respectively. The peak values evaluated by using Eq.~(\ref{EQ04}) are displayed below each subgraph.\label{05SparseAndHol}}
\end{figure}

\begin{figure}[!b]
	\centering
	\includegraphics[width=0.26\textwidth]{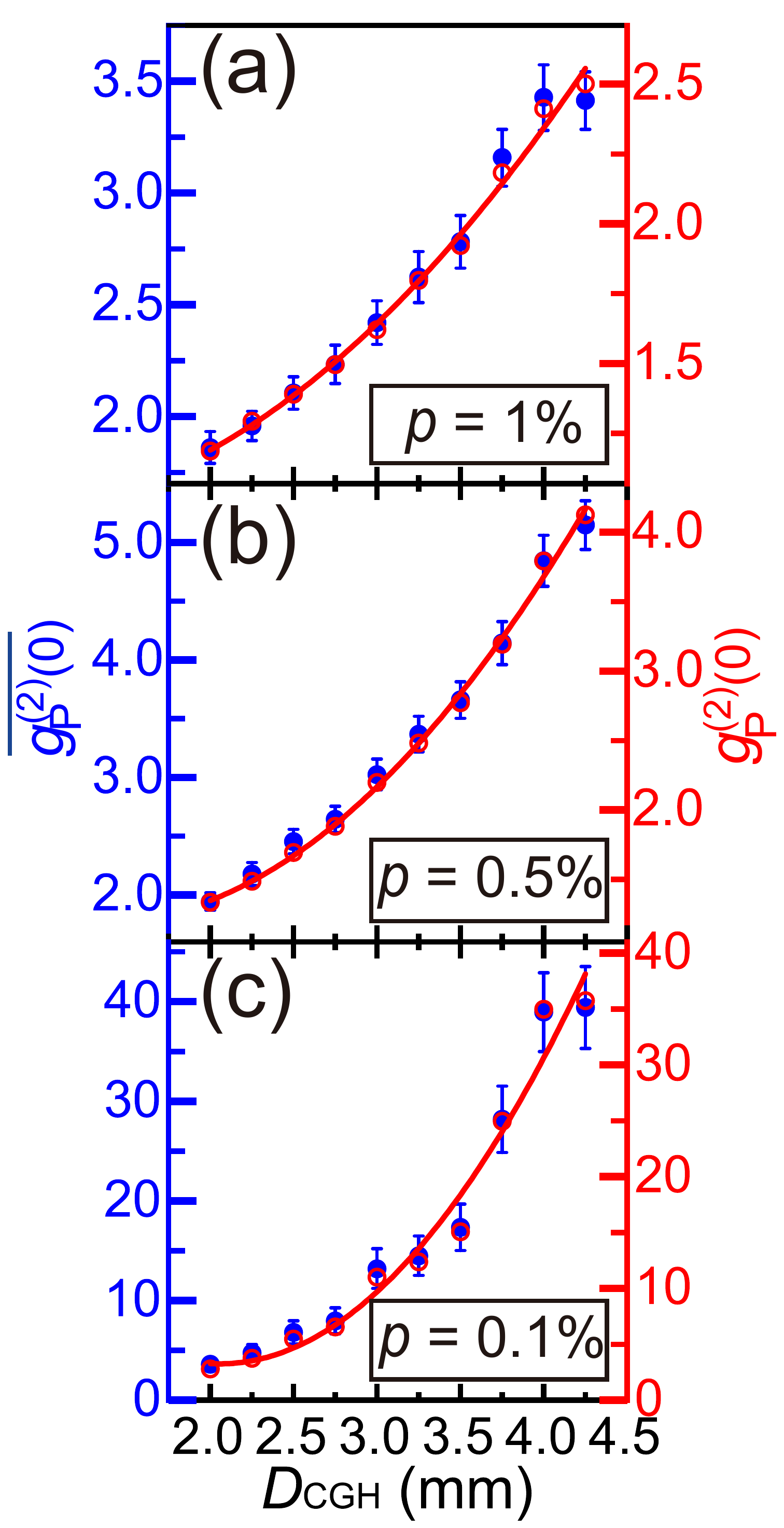}
	\caption{Relationships between $g^{(2)}_{\text{P}}(0)$ and the CGH size $D_{\rm{CGH}}$ for sparse pattens with sparsity ratios $p$ = (a) $1\%$, (b) $0.5\%$, and (c) $0.1\%$ (from top to bottom). The blue solid circles represent the average peak value $\overline{g_{\text{P}}^{(2)}(0)}$ calculated from 100 projection sequences, with error bars (standard deviation) indicated by short blue line segments. The red hollow circles denote $g_{\text{P}}^{(2)}(0)$ derived from the intensity average these 100 frames. The red solid curves represent the theoretical quadratic fits. All other experimental parameters are identical to those in Fig.~\ref{03FunctionByThermal}(d).\label{06FunctionBySparse}}
\end{figure}
Figure~\ref{05SparseAndHol} shows three types of 0-1 binary sparse target patterns and their corresponding holographic projection patterns. The target patterns in Figs.~\ref{05SparseAndHol}(a), \ref{05SparseAndHol}(b), and \ref{05SparseAndHol}(c) were generated from an all-zeros matrix by randomly inserting 1 with proportion $p$ of 1$\%$, 0.5$\%$ and 0.1$\%$, respectively. Given that the GS-CGH projection results for sparse target patterns exhibit superior quality compared to speckle cases at single-frame mode, 100 independent restart GS projection results were experimentally captured for each individual target pattern. Figures~\ref{05SparseAndHol}(d), \ref{05SparseAndHol}(e), and \ref{05SparseAndHol}(f) display one of the 100 projection results corresponding to target patterns in Figs.~\ref{05SparseAndHol}(a), \ref{05SparseAndHol}(b), and \ref{05SparseAndHol}(c), respectively. Furthermore, figures~\ref{05SparseAndHol}(g), \ref{05SparseAndHol}(h), and \ref{05SparseAndHol}(i) present the intensity averages of these 100 projection results for the respective target pattern in Figs.~\ref{05SparseAndHol}(a), \ref{05SparseAndHol}(b), and \ref{05SparseAndHol}(c), respectively. In this experiment, the effective diameter $D_{\rm{CGH}}$  of the CGH was 4 mm. Based on the evaluation by using Eq.~(\ref{EQ04}), the $g_{\text{T}}^{(2)}(0)$ for the target patterns in Figs.~\ref{05SparseAndHol}(a), \ref{05SparseAndHol}(b), and \ref{05SparseAndHol}(c) are 100.06, 199.81 and 992.98, respectively. These results align closely with the theoretical value $g^{(2)}_{\text{T}|p}(0)=1/p$. For the 100 independent restart GS projections results, the average peak values $\overline{g_{\text{P}}^{(2)}(0)}$ and standard deviation are $3.43\pm0.15$, $4.84\pm0.22$ and $38.95\pm3.96$, corresponding to the target patterns in Figs.~\ref{05SparseAndHol}(a), \ref{05SparseAndHol}(b) and \ref{05SparseAndHol}(c), respectively. Additionally, the ${g_{\text{P}}^{(2)}(0)}$ of the projection patterns in Figs.~\ref{05SparseAndHol}(g), \ref{05SparseAndHol}(h), and \ref{05SparseAndHol}(i), obtained by intensity averaging the 100 projection results, decrease to 2.41, 3.79 and 34.96, respectively. It is evident that the peak value of a single-frame projection exceeds that of multi-frame intensity averaging. However, the performance gap in peak values between these two types of projection modes gradually narrows as the sparsity of the target pattern increases. Notably, figures~\ref{05SparseAndHol}(f) and \ref{05SparseAndHol}(i) appear very similar because the holographic reconstruction performs well when the target pattern is extremely sparse ($p$ = $0.1\%$).

\begin{figure}[!b]
	\centering
	\includegraphics[width=0.49\textwidth]{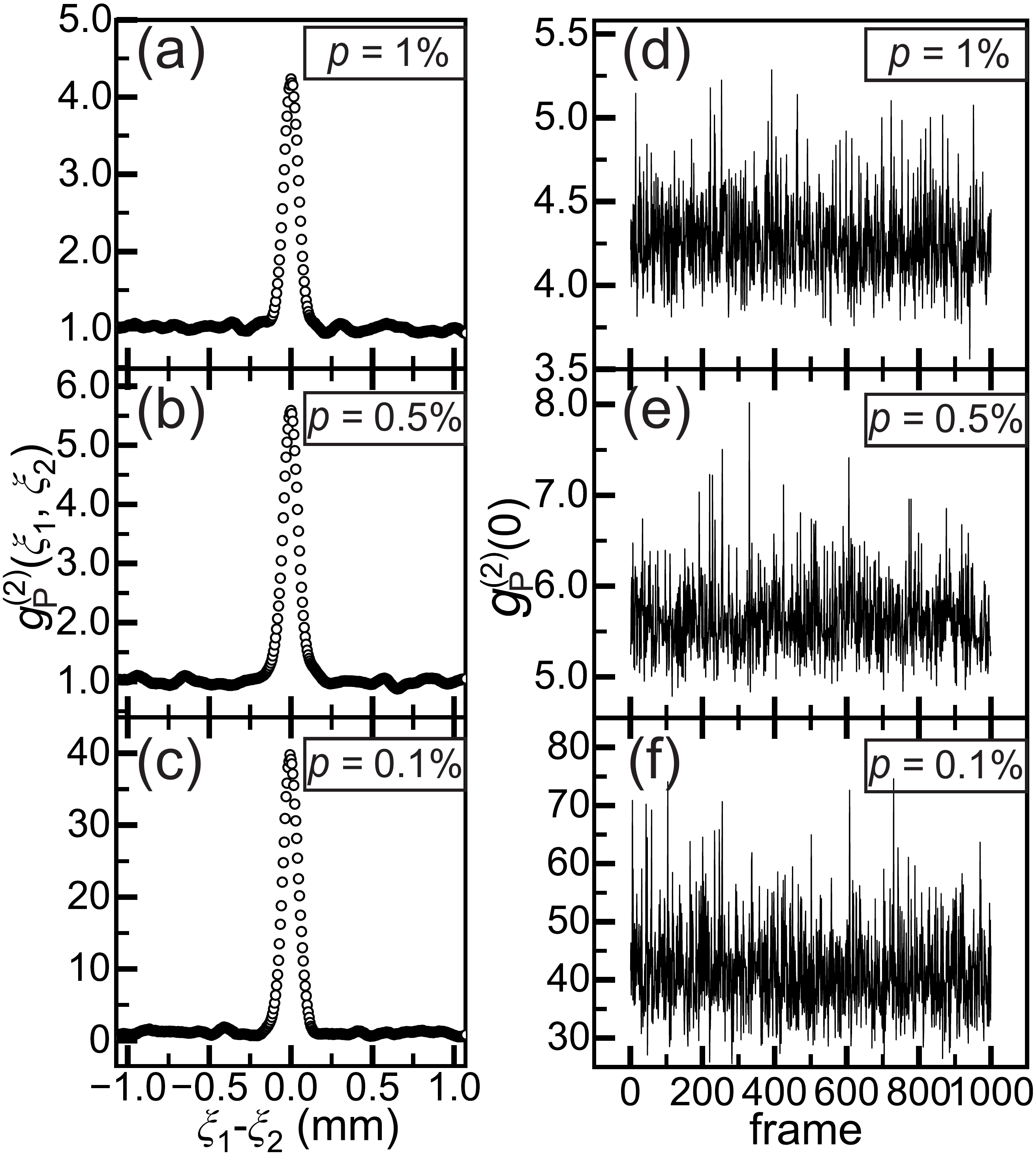}
	\caption{Comparison of two measurement schemes for $g_{\text{P}}^{(2)}(0)$: intensity correlation using multi-frame patterns (a)-(c) and statistical analysis using single-frame patterns (d)-(f). The second-order bunching effect measured via intensity correlation from 10,000 projected frames for sparsity ratios $p$ = (a) $1\%$, (b) $0.5\%$, and (c) $0.1\%$ (from top to bottom). Dynamic curves of $g_{\text{P}}^{(2)}(0)$ measured by Eq.~(\ref{EQ04}) derived from the first 1000 single-frame holographic projections for $p$ = (d) $1\%$, (e) $0.5\%$, and (f) $0.1\%$ (from top to bottom).\label{07Sparseg2bySingleorMulti}}
\end{figure}

To demonstrate the active control of $g^{(2)}_{\text{P}}(0)$ via the numerical aperture of holographic projection, figures~\ref{06FunctionBySparse}(a), \ref{06FunctionBySparse}(b), and \ref{06FunctionBySparse}(c) show the relationships between ${g_{\text{P}}^{(2)}(0)}$ and the CGH size $D_{\rm{CGH}}$ for sparsity ratios $p$ = $1\%$, $0.5\%$, and $0.1\%$, respectively. Expect for the sampling quantity, which was changed from 1,000 to 100, all other experimental parameters remain identical to those in Fig.~\ref{03FunctionByThermal}(d). The blue solid circles represent the average peak value $\overline{g_{\text{P}}^{(2)}(0)}$ calculated from 100 projection sequences for a single CGH size, with error bars (standard deviation) depicted as blue short line segments. The red hollow circles denote $g_{\text{P}}^{(2)}(0)$ obtained from the intensity average of these 100 projected frames, while the red solid curves represent the theoretical fits by a quadratic function. Owing to the pronounced effect of environmental noise at small $D_{\rm{CGH}}$, fitting with Eq.~\eqref{app018} yields unphysical negative results. Thus, a quadratic function is adopted for preliminary fitting in this regime. Overall, a larger CGH size yields finer bright granules and a sparser intensity distribution with enhanced noise immunity, resulting in increased $g_{\text{P}}^{(2)}(0)$. Moreover, $g_{\text{P}}^{(2)}(0)$ obtained under single-frame and multi-frame averaging modes remains strongly linearly correlated, with PCC consistently exceeding 0.998 across three cases in Figs.~\ref{06FunctionBySparse}.

To further identify $g_{\text{P}}^{(2)}(0)$ in holographic projections using sparse patterns, figure~\ref{07Sparseg2bySingleorMulti} compares measurement results obtained via the intensity correlation method and the single-frame statistical analysis method. Figures~\ref{07Sparseg2bySingleorMulti}(a), \ref{07Sparseg2bySingleorMulti}(b), and \ref{07Sparseg2bySingleorMulti}(c) show the experimental results for the bunching effect derived from the 10,000-frame intensity correlation of projection pattens generated from sparse patterns with sparsity ratios $p$ = $1\%$, $0.5\%$, and $0.1\%$, respectively. As the sparsity increases, $g_{\text{P}}^{(2)}(0)$ rises from 4.23 (for $p$ = $1\%$) to 5.59 (for $p$ = $0.5\%$), and further increases to 39.77 (for $p$ = $0.1\%$). Furthermore, dynamic curves of $g_{\text{P}}^{(2)}(0)$ evaluated using only the first 1,000 single-frame projection results are presented in Figs.~\ref{07Sparseg2bySingleorMulti}(d), \ref{07Sparseg2bySingleorMulti}(e), and \ref{07Sparseg2bySingleorMulti}(f), corresponding to $p$ = $1\%$, $0.5\%$, and $0.1\%$, respectively. Here, the average peak values $\overline{g_{\text{P}}^{(2)}(0)}$ and standard deviation calculated from the 10,000-frame patterns for $p$ = $1\%$, $0.5\%$, and $0.1\%$ are $4.24\pm0.25$, $5.58\pm0.41$ and $40.11\pm7.57$, respectively. These results demonstrate that the single-frame evaluation method based on Eq.~(\ref{EQ04}) is effective and can serve as a rapid tool for preliminary assessment. In summary, holographic projection exhibits remarkably strong super-bunching characteristics, particularly when employing sparse target patterns.

\section{Discussion} 
Holographic projection combines two optical systems: imaging and diffraction. From an imaging perspective, if the degree of the second-order coherence function $g^{(2)}_{\text{T}}(0)$ of the target pattern on the object plane exceeds 2, and the spatial resolution of the imaging system reaches a certain threshold, the dynamic intensity on the imaging plane will exhibit a super-bunching effect, as described by Eq.~(\ref{EQ03}). This phenomenon is confirmed by the experimental results represented by red hollow circles in Figs.~\ref{03FunctionByThermal}(c), \ref{03FunctionByThermal}(f), and \ref{06FunctionBySparse}(a)-\ref{06FunctionBySparse}(c), which are derived from the noise suppressed intensity average of multi-frame holographic projection patterns. This behavior is analogous to point-to-spot optical imaging. Notably, improving the spatial resolution of the system facilitates the transfer of the degree $g^{(2)}_{\text{T}}(0)$ of target pattern to the projection plane with minimal loss.

Although holographic reconstruction under a single realization using the GS algorithm can be regarded as a coherent system, the averaging over multiple patterns (analogous to a time-averaged measurement) derived from independent restart GS algorithm constitutes an incoherent system (see Appendix \ref{appendixA} for details). In this work, our theoretical analysis assumes an incoherent model, as it exclusively fits $g^{(2)}_{\text{P}}(0)$ obtained from the averaged intensity patterns generated by these multiple independent GS iterations.

Furthermore, it is well established that the coherent speckle noise arises when a laser beam reflects from a rough surface~\cite{Goodman2007Speckle,Goldfischer1965Far,Dainty2013Laser,Gatti2008ThreeTheory,Magatti2009Three}. For instance, the experimental speckle as shown in Fig.~\ref{01SchemeAndSpeckle}(b) originates from far-field diffraction of coherent light modulated by a random phase. In our work, holographic projection corresponds to the far-field diffraction of a GS-CGH under coherent illumination. Consequently, while such projection noise is unavoidable but it is profitable to enhancing the peak value of the projection pattern.

\begin{figure}[!t]
	\centering
	\includegraphics[width=0.48\textwidth]{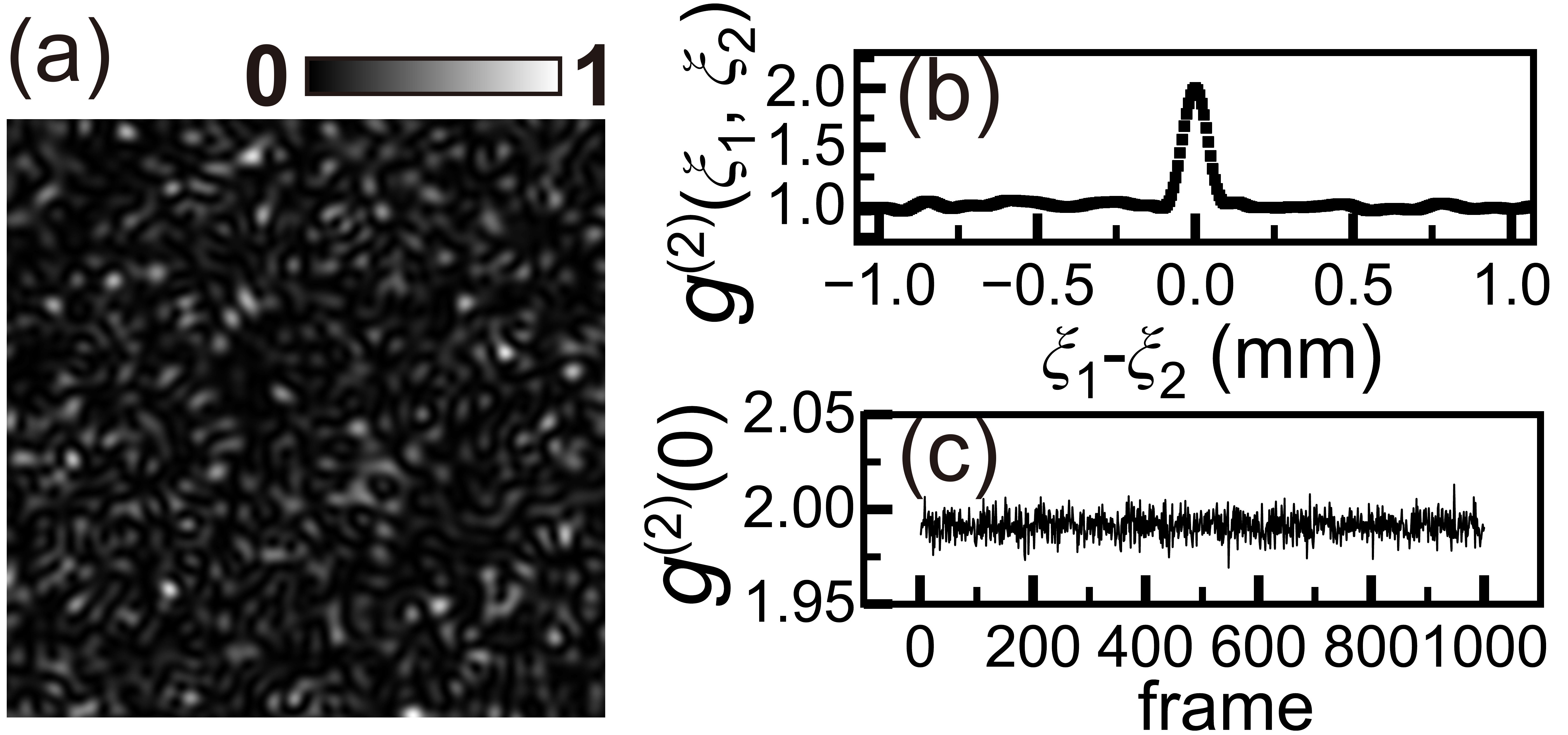}
	\caption{(a) Noise speckle in holographic projection using a uniform target pattern defined by an all-ones matrix. (b) The second-order bunching effect measured via 10,000 frames intensity correlation of the the holographic noise speckle patterns. (c) Dynamic curves of $g^{(2)}(0)$ derived from the first 1,000 holographic noise speckle patterns.\label{08ALLone}}
\end{figure}

To investigate the influence of coherent speckle noise on $g^{(2)}(0)$ in a noise-free environment, we numerically simulated holographic projection using the independent restart GS algorithm with a uniform target pattern defined by an all-ones matrix (no displayed here). Figure~\ref{08ALLone}(a) presents one of the 10,000 holographic projection results generated by this algorithm. In this simulation, the effective diameter $D_{\rm{CGH}}$ of the CGH is 4 mm, while all other parameters remain identical to those in Fig.~\ref{02ThermalAndHol}(b). Figure~\ref{08ALLone}(b) displays the second-order spatial correlation function calculated via multi-frame intensity correlation of these 10,000 projection patterns. The coherent speckle noise exhibits a bunching effect with a full width at half maximum (FWHM) of 89.1 $\mu m$ and a peak value of 2.00. Additionally, figure~\ref{08ALLone}(c) shows the single-frame measured dynamic curves of $g^{(2)}(0)$ derived from the first 1,000 projection results. The average peak value $\overline{g^{(2)}(0)}$ and standard deviation across all 10,000 projection patterns is $1.99\pm0.01$. It is worth noting that the FWHM of the bunching peak represents the average size of the speckle grains, while $g^{(2)}(0)$ is governed by the PDF of the speckle light intensity. Moreover, the dynamic curves in Fig.~\ref{08ALLone}(c) demonstrates that each of the 1,000 noise patterns exhibits nearly identical statistically behavior. Consequently, given the similarity between the bunching effect shown in Fig.~\ref{04Thermalg2bySingleorMulti}(a) and the dynamic curve of $g^{(2)}(0)$ shown in Fig.~\ref{04Thermalg2bySingleorMulti}(b) originated from the Rayleigh speckle, we conclude that these two types of speckle (Rayleigh speckle and coherent speckle noise) share identical physical properties. Beyond the target pattern itself, the coherent speckle noise also modulates the statistical distribution of the projected light intensity, thereby increasing the value of $g_{\text{P}}^{(2)}(0)$. 

Consequently, $g_{\text{P}}^{(2)}(0)$ of the intensity pattern in holographic projection can be effectively controlled through three categories: the degree $g^{(2)}_{\text{T}}(0)$ and coherent length of the target pattern, the physical size of the CGH in the projection system, and the number of frames used for intensity averaging. In computational holography, a CGH can be either amplitude-only or phase-only (adopted in this paper). When combined with Otsu’s threshold method~\cite{OtsuAuto1975AThreshold,GuoArXiv2026Generalized}, an amplitude-only CGH can be binarized and subsequently employed to control $g_{\text{P}}^{(2)}(0)$ using a high‑speed DMD.

Unlike thermal-light speckle, holographic projection enables dynamic modulation of the bunching peak through engineered target patterns. This approach not only simplifies the experimental setup to lensless but also permits high-frame-rate modulation, unlocking capabilities inaccessible to thermal sources. To demonstrate its practical utility in imaging, we recently implemented ghost imaging using this active control strategy~\cite{LimingOE2026Integrating,GuoArXiv2026Generalized}.

\section{Conclusion} 
We theoretically derive analytical formulas for the normalized second-order correlation functions on the holographic projection plane using statistical theory, thereby enabling the theoretical prediction of the bunching peak value. Furthermore, an efficient evaluation method for the bunching peak value is proposed and verified using a single-frame intensity pattern. Experimentally, both Rayleigh speckle and sparse patterns generated via GS-CGH are employed to validate the theoretical results and demonstrate active control over the peak value of the HBT effect. Notably, a super-bunching effect with $g^{(2)}(0)=39.77$ is realized via holographic projection. Additionally, we find that the coherent speckle noise contributes to enhancing $g_{\text{P}}^{(2)}(0)$ of the projection pattern. Finally, we outline several key approaches for controlling $g_{\text{P}}^{(2)}(0)$ within holographic projection schemes employing coherent illumination.

\begin{acknowledgments}
This work was supported by the NSFC (Grants No: 62105188 and 12175127), the Natural Science Foundation of Shandong Province, China (Grant No: ZR2025MS38), and the Scientific Innovation Project for Young Scientists in Shandong Provincial Universities (Grant No: 2024KJG011).
\end{acknowledgments}
\appendix
\section{Intensity expression for holographic reconstruction with noise suppression}\label{appendixA}
\renewcommand{\theequation}{A\arabic{equation}}
\setcounter{equation}{0}
Although holographic projection is distinct from lens-imaging, they share several key commonalities. Notably, when both schemes operate with the same numerical aperture, they yield identical spatial resolutions. In this work, the numerical aperture of holographic projection is defined as the ratio of the CGH radius to the diffraction distance. Given the finite numerical aperture of holographic projection, what is the exact analytical intensity relationship between the target pattern and the reconstructed output?

For CGH generated by the conventional GS algorithm~\cite{Gerchberg1976A}, coherent noise manifests as speckle in the reconstruction pattern, preventing the establishment of a deterministic analytical relationship. For a single target pattern, we generate a series of CGHs using multiple independent restart GS algorithm with different random initial phases. After ensemble averaging over light intensity of a series of reconstruction results, the coherent speckle noise in the holographic multi-frame average mode ($\mathrm{Hol_M}$) is effectively suppressed. In this case, how should the intensity relationship be analytically expressed between $\mathrm{Hol_M}$ and the target pattern.

In the lens imaging schemes, two fundamental frameworks are the coherent imaging model ($\mathrm{Lens_C}$)
 \begin{equation}\label{appA01}
	I_{\mathrm{Lens_C}}(\xi) = \left| \int_{-\infty}^{\infty} \sqrt{I_{\text{T}}(x)} \, h_\text{C}(\xi - x) \, dx \right|^2\\
\end{equation}
and the incoherent imaging model ($\mathrm{Lens_{Inc}}$)
\begin{equation}\label{appA02}
	I_{\mathrm{Lens_{Inc}}}(\xi) =  \int_{-\infty}^{\infty} I_{\text{T}}(x) \, h_{\text{Inc}}(\xi - x) \, dx, \\
\end{equation}
where $I_{\text{T}}(x)$ denotes the intensity distributions at the object plane, $h_{\text{Inc}}(x) = \left| h_\text{C}(x) \right|^2$ represents the point spread function of $\mathrm{Lens_{Inc}}$. Intuitively, one might adopt $\mathrm{Lens_{C}}$ (shown by the Eq.~\ref{appA01}) to describe $\mathrm{Hol_M}$, primarily because holographic reconstruction is inherently a coherent process. Particularly, the source is a laser and no other incoherent modulators are present. However, the coherent speckle noise inherent in the holographic single-frame mode ($\mathrm{Hol_S}$) indicates that the resulting projection pattern is neither strictly coherent nor fully incoherent, but rather constitutes a partially coherent system.

We adopt $\mathrm{Lens_{Inc}}$ (shown by the Eq.~\ref{appA02}) to describe the intensity distribution of $\mathrm{Hol_M}$ for the following three reasons. First, the primary objective of CGH diffraction is to reconstruct the target intensity distribution $I_{\text{T}}(x)$ rather than synthesize its complex amplitude. Although direct Fourier transformation theoretically yields a perfect reconstruction under ideal, noise-free conditions, practical imperfections (coherent wavefront distortion, phase errors and pixel gaps of spatial light modulator) inevitably introduce speckle noise~\cite{Liu2022Double}. Second, inserting a rotating diffuser into a coherent imaging system converts it into an incoherent one via multiframe ensemble averaging. Notably, this process is physically analogous to $\mathrm{Hol_M}$, which emerges from the superposition of statistically independent realizations of multiple $\mathrm{Hol_S}$. Both systems rely on temporally independent, spatially random phases to decorrelate speckle, which is then suppressed through temporal ensemble averaging. Third, the intensity distribution of $\mathrm{Lens_{Inc}}$ and $\mathrm{Hol_M}$ yield essentially equivalent spatial resolution when configured with the same numerical aperture.

To quantitatively validate the analytical description of $\mathrm{Hol_M}$ via $\mathrm{Lens_{Inc}}$, we present a brief comparison of numerical simulation results between holographic projection and single-lens imaging across a broad range of diffraction apertures $D$. In the numerical simulation, the single-lens imaging adopts a $2f$--$2f$ configuration with the diameter $D$ and the focal length $f = 0.3\ \mathrm{m}$ of the lens. For the Fresnel holography projection, the CGH size is set to $D$ with a fixed diffraction distance of $2f$. Under this configuration, both schemes share identical numerical apertures for the same $D$. Here, the intensity results originated from $\mathrm{Hol_M}$ and $\mathrm{Lens_{Inc}}$ were obtained by ensemble averaging over 1000 independent realizations.

\begin{figure}[!b]
	\centering
	\includegraphics[width=0.4\textwidth]{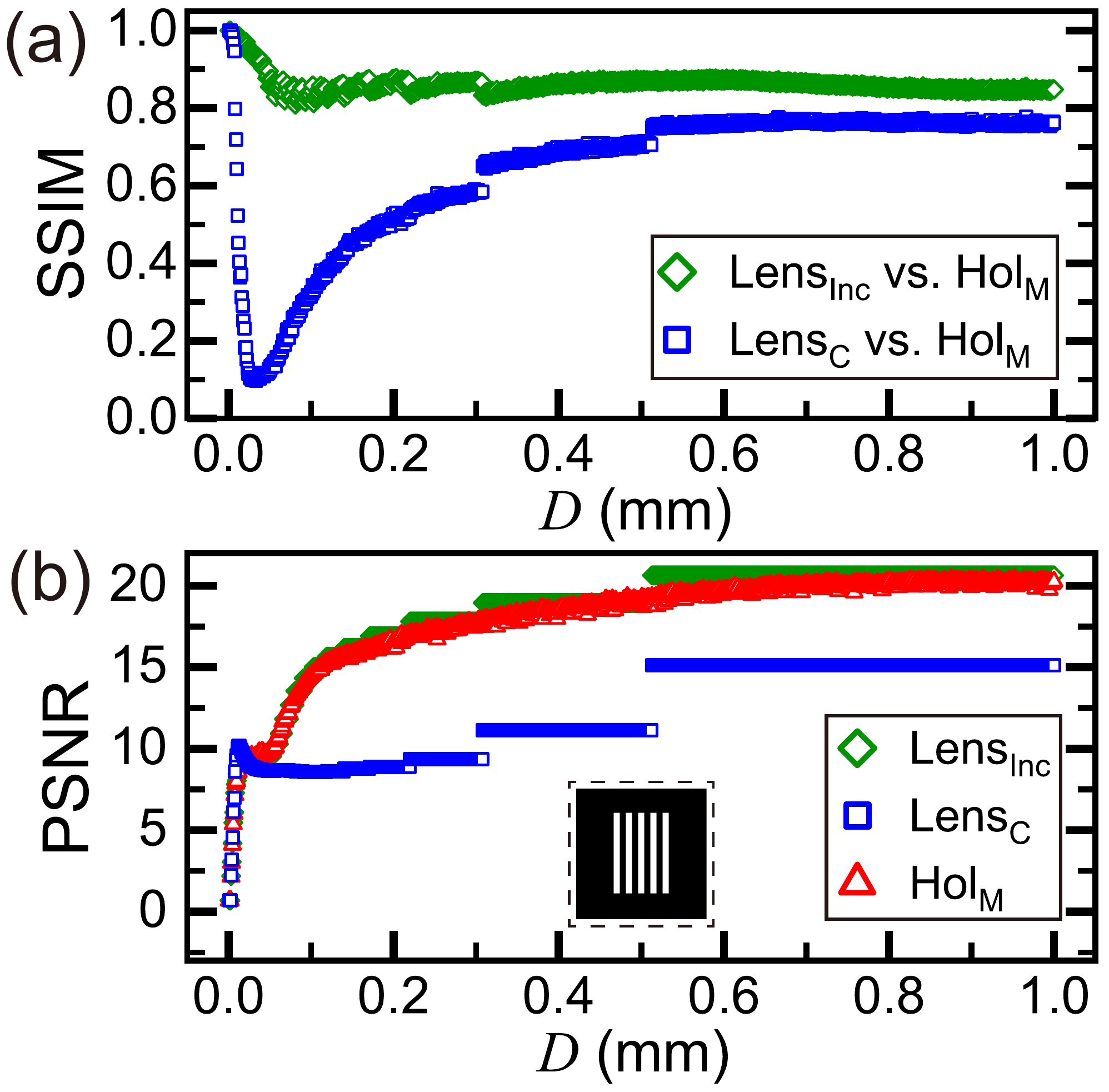}
	\caption{Numerical comparison of holographic projection with multi-frame averaging ($\mathrm{Hol_M}$) and single-lens imaging under coherent ($\mathrm{Lens_C}$) and incoherent ($\mathrm{Lens_{Inc}}$) models at identical numerical apertures (NAs), achieved via matching diffraction apertures ($D$). (a) Evolution of the structural similarity (SSIM) as a function of the diffraction aperture $D$. Green diamonds denote comparisons between $\mathrm{Lens_{Inc}}$ versus $\mathrm{Hol_M}$, while blue squares denote comparisons between $\mathrm{Lens_C}$ versus $\mathrm{Hol_M}$. (b) Peak signal-to-noise ratio (PSNR) for holographic reconstruction ($\mathrm{Hol_M}$, red triangles), lens coherent imaging ($\mathrm{Lens_C}$, blue squares), and lens incoherent imaging ($\mathrm{Lens_{Inc}}$, green diamonds), all evaluated against the ground-truth test target (shown in the inset with a black dashed box), a binary (0--1) striped object comprising five equally spaced slits.\label{09CorINC}}
\end{figure}

Figure~\ref{09CorINC} presents the structural similarity (SSIM)~\cite{ZhouProcessing2024Image} and the peak signal-to-noise ratio (PSNR)~\cite{LI2021127749} as quantitative metrics for assessing the consistency between $\mathrm{Lens_{Inc}}$ and $\mathrm{Hol_M}$. In Fig.~\ref{09CorINC}(a), two SSIM curves are presented: green diamonds correspond to comparisons between $\mathrm{Lens_{Inc}}$ and $\mathrm{Hol_M}$, while blue squares correspond to comparisons between $\mathrm{Lens_C}$ and $\mathrm{Hol_M}$. All results are computed at identical aperture values $D$. For the $\mathrm{Lens_{Inc}}$ versus $\mathrm{Hol_M}$ curve, the SSIM remains above 0.81 across the full range of $D$, stabilizing at $0.86 \pm 0.01$ for $D > 0.5\ \mathrm{mm}$. In contrast, the curve of $\mathrm{Lens_C}$ versus $\mathrm{Hol_M}$ exhibits severe degradation at small apertures ($D < 0.3\ \mathrm{mm}$), with only partial recovery to $0.76 \pm 0.01$ for $D > 0.5\ \mathrm{mm}$. Figure~\ref{09CorINC}(b) displays three PSNR curves that compare the intensity distributions obtained from the imaging and holographic projection schemes with the ground-truth target pattern (shown in the inset with a black dashed box). The target is a binary (0--1) striped object consisting of five equally spaced slits. In Fig.~\ref{09CorINC}(b), the green diamonds, blue squares, and red triangles correspond to the measurement results of $\mathrm{Lens_{Inc}}$, $\mathrm{Lens_C}$, and $\mathrm{Hol_M}$, respectively. All three methods exhibit improved reconstruction quality with increasing $D$. Notably, the curves originating from $\mathrm{Lens_{Inc}}$ and $\mathrm{Hol_M}$ exhibit negligible deviation, indicating near-identical performance. In summary, both metrics (SSIM and PSNR) confirm that employing $\mathrm{Lens_{Inc}}$ (Eq.~\ref{appA02}) to analytically characterize $\mathrm{Hol_M}$ is well justified.

\section{Expected value of a random variable}\label{appendixB}
\renewcommand{\theequation}{B\arabic{equation}}
\setcounter{equation}{0}

Type one: $X$ is a discrete random variable.
\begin{equation}\label{app001}
	E( X ) = \sum\limits_{i = 1}^\infty  {{X_i}{p_i({X = {X_i}})}},
\end{equation}
where $p_i$ is the probability that $X$ equals $X_i$.

Type two: $X$ is a continuous random variable.
\begin{equation}\label{app002}
	E( X ) = \int_{ - \infty }^{ + \infty } {x {f_X}(x){\rm{d}}x},
\end{equation}
where ${f_X}(x)$ is the PDF of the random variable $X$.

Type three: Let $Z = g_1(X)$ be a random variable as a function of the random variable $X$, where $g_1(X)$ is known. Depending on the type of $X$, the expected value of $Z$ can be expressed as~\cite{Leon2017Probability}:
\begin{equation}\label{app003}
	\begin{split}	
		E( Z ) =  \sum\limits_{i = 1}^\infty  {g_1(X)p_i({X = {X_i}})},\\
	\end{split}
\end{equation}
or
\begin{equation}\label{app004}
	\begin{split}	
		E( Z )= \int_{ - \infty }^{ + \infty } { g_1(x){f_X}(x){\rm{d}}x}.\\
	\end{split}
\end{equation}

Type four: Let $Z = g_2( X, Y )$ be a random variable as a function of two random variables $X$ and $Y$, where the function $g_2(X,Y)$ is known. The expected value of $Z$ can be expressed as~\cite{Leon2017Probability}:
\begin{equation}\label{app005}
	\begin{split}	
		&E(Z)= \sum\limits_{i = 1}^\infty \sum\limits_{j = 1}^\infty  {g_2(X,Y)p_i( {X = {X_i}} )p_j( {Y = {Y_j}})},\\
	\end{split}
\end{equation}
or
\begin{equation}\label{app006}
	\begin{split}	
		E( Z )= \int_{ - \infty }^{ + \infty } g_2(X,Y){f_{(X,Y)}}( x,y ){\rm{d}}x {\rm{d}}y.\\
	\end{split}
\end{equation}
Here, ${f_{(X,Y)}}(x,y)$ is the joint PDF of random variables $X$ and $Y$.

\section{Expected values of dynamic speckle and its power transformation}\label{appendixC}
\renewcommand{\theequation}{C\arabic{equation}}
\setcounter{equation}{0}

For dynamic speckle pattern ${I_\text{T}}( {x,t} )$, the PDF~\cite{Goodman1976Some} is
\begin{equation}\label{app007}
	\begin{split}	
		f_{I_\text{T}}(i_\text{T})=1/{\bar{I}_{\text{T}}}\cdot \text{exp}(-{i_\text{T}}/{\bar{I}_{\text{T}}}),
	\end{split}
\end{equation}
where
$i_\text{T}>0$ and $\bar{I}_{\text{T}}=\left \langle {I_\text{T}}( {x,t} ) \right \rangle_t$ denotes the expected value over $t$ at any point on the target plane. Based on Eqs.~(\ref{app004}) and (\ref{app007}), the expected value of the $\kappa$th power transformation of thermal speckle can be expressed as:
\begin{equation}\label{app008}
	\begin{split}	
		\langle I_{\text{T}}^\kappa(x,t) \rangle_t=\int_{0}^{+\infty}i_\text{T}^\kappa f_{I_\text{T}}(i_\text{T})di_\text{T}=g^{(\kappa)}_{\text{ther}}(0)\bar{I}_{\text{T}}^\kappa,
	\end{split}
\end{equation}
where $g^{(\kappa)}_{\text{ther}}(0)$ denotes the degree of the $\kappa$th-order coherence function of dynamic thermal speckle. For integer values of $\kappa$, this degree satisfies $g^{(\kappa)}_{\text{ther}}(0)= \kappa!$. Notably, this relationship also holds for $\kappa=0.5$. Based on the generalized factorial~\cite{Bhargava2000The}, the degree of the coherence function of thermal speckle at $\kappa=0.5$ satisfies $g^{(0.5)}_{\text{ther}}(0)=\frac{1}{2}!=\frac{\sqrt{\pi}}{2}$.

\section{The second-order spatial correlation function of intensity pattern in the holographic projection multi-frame average mode}\label{appendixD}
\renewcommand{\theequation}{D\arabic{equation}}
\setcounter{equation}{0}  

Based on Eqs.~(\ref{EQ01}), (\ref{app004}), (\ref{app007}), and (\ref{app008}), the expected value of the intensity distribution of the projection pattern at position $\xi$ can be expressed as:
\begin{equation}\label{app009}
	\begin{split}	
		\left \langle{I_\text{P}}( {\xi ,t} )   \right \rangle_t=\sum\limits_{i = 1}^\mathcal{N} g^{(\kappa)}_{\text{ther}}(0)\bar{I}_{\text{T}}^\kappa \cdot {{\rm{somb}}^2}\left[ {k{N_{\rm{A}}}( {\xi  - {x_i }} )} \right].\\
	\end{split}
\end{equation}

To derive the expected value of the product of the projection patterns at two detection points, we extend the formulation from a special case to the general scenario. Assuming that effective intensity correlation exists only between adjacent points (within a range of $\pm 1$) on the target pattern, the product of the light intensities can be expressed as:

\begin{equation}\label{app010}
	\begin{split}	
		&{I_\text{P}}( {\xi_1 ,t} ){I_\text{P}}( {\xi_2 ,t} )\\
		&=\sum\limits_{i  = 1}^\mathcal{N} {{I_\text{T}^\kappa}( {{x_i },t} )\cdot{{\rm{somb}}^2}\left[ {k{N_{\rm{A}}}( {\xi_1  - {x_i }} )} \right]}\\
		&\times    \sum\limits_{j  = 1}^\mathcal{N} {{I_\text{T}^\kappa}( {{x_j },t} )\cdot{{\rm{somb}}^2}\left[ {k{N_{\rm{A}}}( {\xi_2  - {x_j }} )} \right]}\\
		&=\mathcal{P}_1+\mathcal{P}_2,
	\end{split}
\end{equation}
where $\mathcal{P}_1$ corresponds to the case of intensity non-correlation terms ($i\ne j,j\pm 1$):
\begin{equation}\label{app011}
	\begin{split}	
		\mathcal{P}_1=&{I_\text{P}}( {\xi_1 ,t} ){I_\text{P}}( {\xi_2 ,t} )\cdot \left(1-\delta_{i,j}-\delta_{i,j+1}-\delta_{i,j-1}\right) ,\\
	\end{split}
\end{equation}
and $\mathcal{P}_2$ corresponds to the case of intensity correlation terms ($i= j,j\pm 1$):
\begin{equation}\label{app012}
	\begin{split}	
		\mathcal{P}_2={I_\text{P}}( {\xi_1 ,t} ){I_\text{P}}( {\xi_2 ,t} )\cdot ( \delta_{i,j}+\delta_{i,j+1}+ \delta_{i,j-1} ),\\
	\end{split}
\end{equation}
where $\delta_{i,j} = 1$ for $i=j$, and $\delta_{i,j} = 0$ otherwise. Here, the product ${I_\text{P}}( {\xi_1 ,t} ){I_\text{P}}( {\xi_2 ,t} )$ is decomposed into two components $\mathcal{P}_1$ and $\mathcal{P}_2$, which originate from the interaction of two random variables, ${I_\text{T}^\kappa}\left( {{x_i },t} \right)$ and ${I_\text{T}^\kappa}\left( {{x_j },t} \right)$. In $\mathcal{P}_1$, these two random variables are completely independent, whereas in $\mathcal{P}_2$, they are correlated. Based on Eqs.~(\ref{app006}) and (\ref{app009})-(\ref{app012}), the ensemble average of ${I_\text{P}}( {\xi_1 ,t} ){I_\text{P}}( {\xi_2 ,t} )$ can be expressed as:
\begin{widetext}
	\begin{equation}\label{app013}
		\begin{split}	
			\left \langle {I_\text{P}}( {\xi_1 ,t} ){I_\text{P}}( {\xi_2 ,t} ) \right \rangle_t &=\left \langle \mathcal{P}_1\right \rangle_t +\left \langle \mathcal{P}_2\right \rangle_t\\
			&=(g^{(\kappa)}_{\text{ther}}(0))^2 \bar{I}_{\text{T}}^{2\kappa}   \sum\limits_{i  = 1}^\mathcal{N}   
			{{\rm{somb}}^2}\left[ {k{N_{\rm{A}}}( {\xi_1  - {x_i }} )} \right] 
			\sum\limits_{j  = 1}^\mathcal{N}  
			{{\rm{somb}}^2}\left[ {k{N_{\rm{A}}}( {\xi_2  - {x_j }} )} \right]\\
			&-(g^{(\kappa)}_{\text{ther}}(0))^2 \bar{I}_{\text{T}}^{2\kappa}   \sum\limits_{i  = 1}^\mathcal{N}   
			{{\rm{somb}}^2}\left[ {k{N_{\rm{A}}}( {\xi_1  - {x_i }} )} \right]{{\rm{somb}}^2}\left[ {k{N_{\rm{A}}}( {\xi_2  - {x_i }} )} \right]\\
			&-(g^{(\kappa)}_{\text{ther}}(0))^2 \bar{I}_{\text{T}}^{2\kappa}   \sum\limits_{i  = 1}^\mathcal{N}   
			{{\rm{somb}}^2}\left[ {k{N_{\rm{A}}}( {\xi_1  - {x_i }} )} \right]{{\rm{somb}}^2}\left[ {k{N_{\rm{A}}}( {\xi_2  - {x_i+\Delta }} )} \right]\\
			&-(g^{(\kappa)}_{\text{ther}}(0))^2 \bar{I}_{\text{T}}^{2\kappa}   \sum\limits_{i  = 1}^\mathcal{N}   
			{{\rm{somb}}^2}\left[ {k{N_{\rm{A}}}( {\xi_1  - {x_i }} )} \right]{{\rm{somb}}^2}\left[ {k{N_{\rm{A}}}( {\xi_2  - {x_i-\Delta }} )} \right]\\
			&+(g^{(\kappa)}_{\text{ther}}(0))^2 g_{\text{T}}^{\left ( 2 \right ) }  ( 0   )\bar{I}_{\text{T}}^{2\kappa}    \sum\limits_{i  = 1}^\mathcal{N} {{{\rm{somb}}^2}\left[ {k{N_{\rm{A}}}( {\xi_1  - {x_i }} )} \right]}{{{\rm{somb}}^2}\left[ {k{N_{\rm{A}}}( {\xi_2  - {x_i }} )} \right]}, \\
			&+(g^{(\kappa)}_{\text{ther}}(0))^2 g_{\text{T}}^{\left ( 2 \right ) }  ( -\Delta   )\bar{I}_{\text{T}}^{2\kappa}   \sum\limits_{i  = 1}^\mathcal{N}   
			{{\rm{somb}}^2}\left[ {k{N_{\rm{A}}}( {\xi_1  - {x_i }} )} \right]{{\rm{somb}}^2}\left[ {k{N_{\rm{A}}}( {\xi_2  - {x_i+\Delta }} )} \right]\\
			&+(g^{(\kappa)}_{\text{ther}}(0))^2 g_{\text{T}}^{\left ( 2 \right ) }  ( \Delta   )\bar{I}_{\text{T}}^{2\kappa}   \sum\limits_{i  = 1}^\mathcal{N}   
			{{\rm{somb}}^2}\left[ {k{N_{\rm{A}}}( {\xi_1  - {x_i }} )} \right]{{\rm{somb}}^2}\left[ {k{N_{\rm{A}}}( {\xi_2  - {x_i-\Delta }} )} \right],\\
		\end{split}
	\end{equation}
\end{widetext}
where $\Delta = \left | x_i-x_{i+1} \right |$ denotes the distance between adjacent points on the target plane. The second-order intensity correlation of the dynamic light intensity between two points $x_i$ and $x_j$ (or equivalently, between the two random variables ${I_\text{T}^\kappa}( {{x_i },t} )$ and ${I_\text{T}^\kappa}( {{x_j },t} )$) on the target plane is given by Eq.~(\ref{EQ02}). Therefore, based on Eqs.~(\ref{EQ02}), (\ref{app009}), and (\ref{app013}), the general form of the normal second-order spatial correlation function at the projection plane can be expressed as:
\begin{widetext}
	\begin{equation}\label{app015}
		\begin{split}
			g_{\text{P}}^{(2)}(\xi_1, \xi_2)&=\frac{\left \langle {I_\text{P}}( {\xi_1 ,t} ){I_\text{P}}( {\xi_2 ,t} ) \right \rangle_t}{\left \langle{I_\text{P}}({\xi_1 ,t})\right \rangle_t \left \langle{I_\text{P}}({\xi_2 ,t})\right \rangle_t}\\
			&=1+\frac{\sum\limits_{i = 1}^\mathcal{N} 
				\left({{\rm{somb}}^2}[ {k{N_{\rm{A}}}(\xi_1-{x_i})}]
				\sum\limits_{j=1}^{\mathcal{N}}[g_{\text{T}}^{(2)}(x_i-x_j)-1] \cdot
				{{\rm{somb}}^2}[{k{N_{\rm{A}}}\left({\xi_2  - {x_j }} \right)}]\right) }
			{   \left(\sum\limits_{i= 1}^\mathcal{N} {{\rm{somb}}^2}\left[ {k{N_{\rm{A}}}( {\xi_1 - {x_i }} )} \right]\right)^2}.\\
		\end{split}	
	\end{equation}
\end{widetext}
For a continuous system, the counterpart of Eq.~(\ref{app015}) can be expressed as:
\begin{widetext}
	\begin{equation}\label{app016}
		\begin{split}
			g_{\text{P}}^{(2)}(\xi_1, \xi_2)=1+\frac{\int_{-\infty}^{+\infty}\int_{-\infty}^{+\infty}{{{\text{somb}}^{2}}  (kN_{\rm{A}}x) \cdot{{\text{somb}}^{2}}(kN_{\rm{A}}x')\cdot[ g_{\text{T}}^{(2)}(\xi_1-\xi_2-x+x')-1 ]dxdx'}}{{{\left( \int{{{\text{somb}}^{2}}(kN_{\rm{A}}x)dx} \right)}}^{2}  }.\\
		\end{split}	
	\end{equation}
\end{widetext}
Note that, $g_{\text{P}}^{(2)}(\xi_1, \xi_2)$ is independent of the specific positions and depends solely on the separation distance $\xi_1-\xi_2$ between the two points in the detected plane. This implies that the bunching effect of the projection pattern satisfies spatial translation invariance, a necessary condition for ghost imaging schemes~\cite{WuCPB2024High,Goodman1995Introduction}. Moreover, the analytical results derived from Eqs.~(\ref{app015}) and (\ref{app016}) are applicable not only to thermal light speckle and its functionally transformed patterns but also to an arbitrary target pattern exhibiting statistically independent light intensity in both spatial and temporal domains. This includes, for example, the sparse target patterns designed in the experimental verification section (Section~\ref{sec:Sparse}). However, it should be noted that the corresponding expression $g_{\text{T}}^{(2)}$ in Eqs.~(\ref{app015}) and (\ref{app016}) requires modification.

Based on the statistical properties of the 0-1 binary sparse target pattern, the normalized second-order spatial correlation function of the target pattern can be expressed as:
\begin{equation}\label{app017}
	g^{(2)}_{\text{T}|p}(x_i, x_j)=1+[g^{(2)}_{\text{T}|p}(0)-1]\cdot \delta(x_i-x_j),\\
\end{equation}
where $g^{(2)}_{\text{T}|p}(0)=p^{-1}$. Here, the non-zero intensity regions behave as ideal point sources. For a discrete system, combining Eqs.~(\ref{app015}) and (\ref{app017}), the peak value of the HBT effect in the projection imaging plane is given by:
\begin{equation}\label{app018}
	g_{\text{P}}^{(2)}(0)=1+(p^{-1}-1)\cdot\frac{\sum\limits_{i= 1}^\mathcal{N} {{\rm{somb}}^4}\left( {k{N_{\rm{A}}}{x_i}} \right)}{\left(\sum\limits_{i= 1}^\mathcal{N} {{\rm{somb}}^2}\left( {k{N_{\rm{A}}}{x_i}} \right)\right)^2}.\\
\end{equation}
Consequently, a limited $N_{\rm{A}}$ reduces the peak value of the HBT effect in the holographic projection pattern.
\bibliography{HBTbyHolo.bib}
\end{document}